# Daily Stand-Up Meetings
## Start Breaking the Rules

PREPRINT[1]

**Viktoria Stray, University of Oslo, SINTEF**
**Nils Brede Moe, SINTEF**
**Dag I.K. Sjøberg, University of Oslo, SINTEF**

**ABSTRACT. Members of high performing software teams collaborate, exchange information and coordinate their work on a frequent, regular basis. Most teams have the daily stand-up meeting as a central venue for these activities. Although this kind of meeting is one of the most popular agile practices, it has received little attention from researchers. We observed 102 daily stand-ups and interviewed 60 members of 15 teams in five countries. We found that the practice is usually challenging to conduct in a way that benefits the whole team. Many team members have a negative experience from conducting the meeting, which reduces job satisfaction, co-worker trust and well-being. However, the practice can be adjusted and improved to empower teams. In this article, we describe key factors that affect the meeting and propose four recommendations for improving the practice.**

If you work in an agile project, you are probably familiar with some format of the daily stand-up meeting, since it is one of the most popular agile practices [1], [2]. You may see the meeting as a necessary venue for communication with the team, or perhaps you consider it a waste of time. We have been researching daily stand-up meetings for almost a decade and helped software engineering companies find ways to improve their meetings. In this article, we share our findings and provide recommendations on how to conduct daily stand-up meetings.

Agile methods introduced the practice of daily stand-ups to improve communication in software development teams and projects. In these meetings, team members share information and thus become aware of what other members are doing. The intention is that team members will then align their actions to fit the actions of the others, that is, mutual adjustment [3]. Furthermore, meetings that improve access to information fosters employee empowerment [4].

---

[1] (C) IEEE, accepted for publication in IEEE Software.

Despite clear guidelines of how the 15-minute meeting should be conducted [5], we have through our visits to more than 40 companies and several hundred teams found that implementing daily stand-ups in a way that benefits the whole team is quite challenging. So, how can we adjust the practice to boost team performance? Understanding how to conduct these meetings for self-managing software development teams requires more than just examining the team's inner workings. We must also understand the organisational context surrounding the team. Therefore, we have conducted a multiple-case study in four software companies to identify positive and negative aspects of daily stand-up meetings and provide recommendations on how to get more value from the meetings.

## Research Background [may be a sidebar]

The four companies studied belonged to different application domains. ITConsult is a Norwegian IT consulting company, TelSoft is an international telecommunications software company, GlobEng is a global software company providing services for the engineering industry, and NorBank is a Nordic bank and insurance company. Data collection and analysis took place from 2010 to 2018.

To collect different perspectives on the meeting, we interviewed 60 project members with roles at all levels in 15 different teams, see Table 1. The overall topic of the interviews was teamwork and meetings, with a particular focus on daily stand-up meetings. More details on the interview guide and research context are available in [7] and [8]. We observed and documented 102 daily stand-up meetings. Eight of the meetings from TelSoft and NorBank were audiotaped, transcribed and finally coded based on a validated coding scheme for team meeting processes reported in [6]. Figure 1 shows the proportion of words stated in each of the six categories we identified. The number of words can be seen as a proxy for the time spent.

**TABLE 1:** Data sources

| Company | ITConsult | TelSoft | GlobEng | NorBank | Total/avg |
|---|---|---|---|---|---|
| Locations visited | Poland, Norway | Malaysia, Norway | China, Poland, Norway, UK | Norway | 9 sites |
| No. of teams studied | 2 | 3 | 7 | 3 | 15 |
| Avg. team size (min/max) | 9.5 (9/10) | 9.3 (9/10) | 8.3 (5/13) | 14 (13/14) | 10.3 |
| No. of interviews | 15 | 20 | 13 | 12 | 60 |
| No. of stand-up meetings observed | 19 | 39 | 13 | 31 | 102 |
| Avg. stand-up duration (min/max) | 10.8 (4/18) | 15.5 (7/24) | 11.6 (4/18) | 8.9 (4/13) | 11.7 |

**TABLE 2:** Daily stand-up meetings: Benefits and problems

| Main benefits |
|---|
| • Problems are quickly identified, discussed and resolved |
| • Team cohesion and shared commitment is increased |
| • Higher awareness of what other team members are doing |
| • Better coordination of interactions through mutual adjustment |
| • More effective decision making |
| **Main problems** |
| • Information shared is not perceived as relevant, particularly due to diversity in roles, tasks and seniority |
| • Managers or Scrum Masters use the meeting primarily to receive status information |
| • Productivity is reduced because the day is broken into slots |

## Daily stand-ups – what, how, when and for whom?

The interviewees expressed that discussions other than answering the three Scrum questions was most valuable. Furthermore, the value of the stand-ups was affected by meeting facilitation behaviour, the dependencies among tasks and the time of day the meetings were held. In addition, we found that juniors were more satisfied with the meetings than seniors. See Table 2 for a summary of the main benefits and problems of conducting daily stand-ups.

### What is discussed in the meetings?

Traditionally, participants in daily stand-ups are supposed to answer a variant of only these three questions: "What did I do yesterday?" (Q1) "What will I do today?" (Q2) and "Do I see any impediments?" (Q3). We found that the teams addressed far more than the three questions within the 15 minutes. The analysis of the transcribed meetings showed that, on average, only 34% of the meeting was spent on answering the three questions, see Figure 1. The teams spent almost as much time (31%) on elaborating problem issues, discussing possible solutions and making decisions, also known as problem-focused communication [6]. According to Scrum, the daily stand-up should not be used for discussing solutions to obstacles raised. However, in the interviews, problem-focused communication was the most frequently mentioned positive aspect of attending stand-ups. The teams highly valued the ability to have an arena for quick problem-solving, even if the problem-solving sessions were as short as a minute.

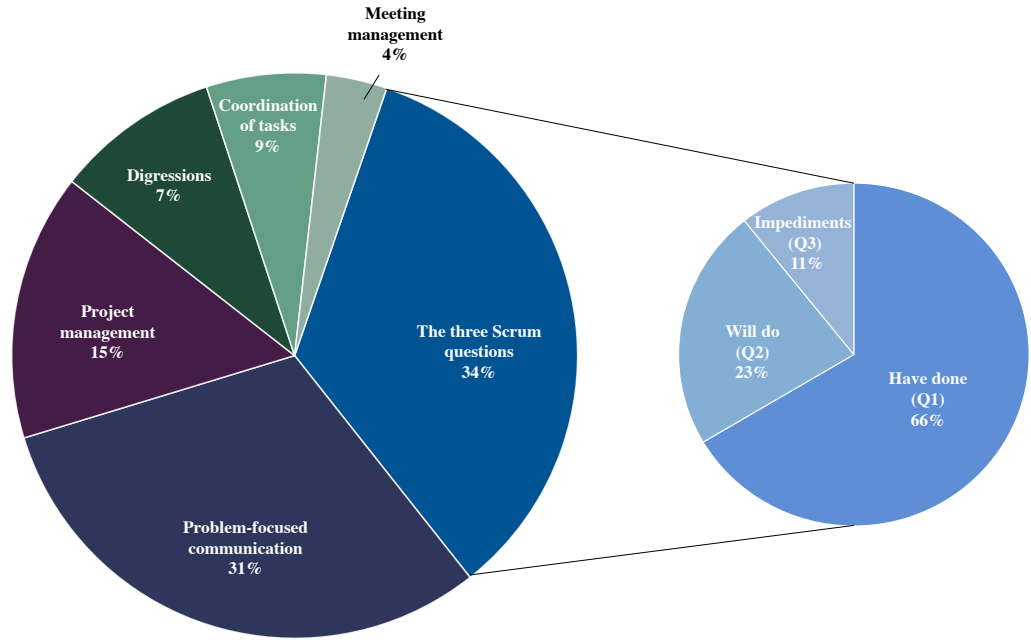

**Figure 1:** Proportion of topics discussed in daily stand-up meetings.

The daily stand-up is characterised by limited time, uncertainty and incomplete information. A reasonable question to ask then is whether it is worthwhile to spend meeting time on decision-making. A successful problem-solving process is often characterised by a thorough definition and analysis of the problem, which may be time-consuming. To understand decision-making in such a context, we use the theory of naturalistic decision-making (NDM) [9]. NDM postulates that experts can make the right decisions under challenging conditions, such as time pressure, uncertainty and vague goals, without having to perform extensive analyses and compare options. They use their experience to recognise problems that they have previously encountered. They form mental simulations of the problem currently being faced and use these simulations to suggest appropriate solutions quickly. Because problem-focused communication is essential for team problem-solving [6], and the meeting is the only daily coordination arena for the whole team, we argue that the meeting is necessary for effective decision-making in agile teams.

Both meeting transcripts and observations showed that the problem-focused communication usually happened after the team members had described what they were going to do (Q2) or had mentioned an obstacle (Q3). Just relaying what they had been doing (Q1) rarely triggered problem-focused communication. Therefore, Q2 and Q3 are clearly most important for making quick decisions. While the interviewees stated to be least happy to spend time on Q1, the teams actually spent the most time on that question. The main reason was that the talking about what had been done often ended up as cumbersome reporting with superfluous details. The challenge of knowing what others need to know tended to result in team members trying to include everything they had worked on independently of its relevance to others. In one team that did not spend much time on Q1, the Scrum Master explained: *"We don't want to know the details of what you did yesterday, but something that is useful for the whole team. We had to change the mindset and find out what to say in the meeting. We practised a lot and improved"*.

**How are the meetings facilitated?**
We observed that if the behaviour of the meeting facilitator made team members address the facilitator, the stand-up often became a status-reporting meeting. If the facilitator managed to make team members talk to each other, the stand-up tended to become a discussion meeting. To illustrate the two types of meetings, we show conversation charts from two observations in TelSoft (Figures 2 and 3). The lines represent an interaction between two participants. Figure 2 shows a reporting meeting where the facilitator controlled the talking by allocating turns for speaking. Through this behaviour, team members naturally directed their response to her and had less interaction with each other. Figure 3 shows a discussion meeting where team members engaged in what the other members were saying. The facilitators who managed to get people to talk to each other were usually involved in the day-to-day work and solving of team tasks, and therefore had less need for receiving status information.

Team members who practiced daily stand-ups as reporting meetings expressed negative attitudes towards these meetings. One member said it felt like having oral exams every day; another said: *"No-one in the team really wants to be at the status meeting"*. When the regular facilitator of such meetings was absent, usually either the team dropped the meeting that day, or they conducted a discussion meeting instead. One developer explained: *"Today the Scrum Master was not there, and we probably had our best daily meeting. Because usually, the meetings are focused on information useful to the Scrum Master, but not for us. He is also more interested in what we have done than what we are going to do."* Other research has found that poor meetings may have a negative effect beyond the meetings themselves, such as reduced job satisfaction, co-worker trust and well-being [4], [11].

When the facilitator had a personal interest in the status of specific tasks, they often gave more attention to team members working on these tasks, leading to an unbalanced contribution from the team members in the meeting. One team member said, *"The Scrum Master chooses the next person to talk based on the tasks that are important to her"*. We observed that people who talked early were given the time they wanted, while the people who talked last was cut short by the Scrum Master because the meeting was approaching the time limit. It is negative for a team if, for example, developers are given more time to speak than testers. To make the team productive, everyone in the team should, over time, speak roughly the same amount of time [10].

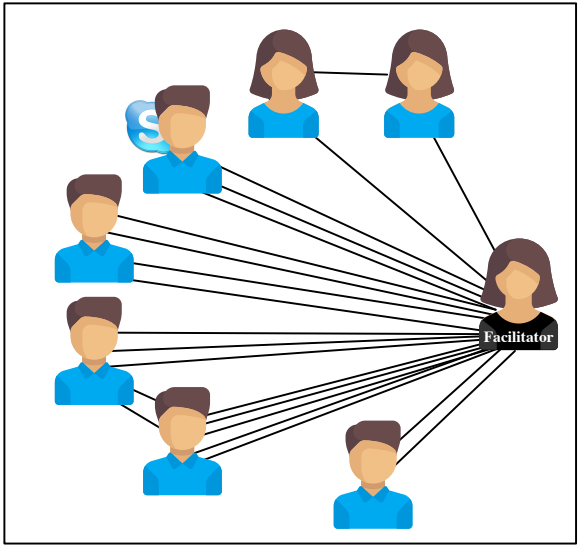
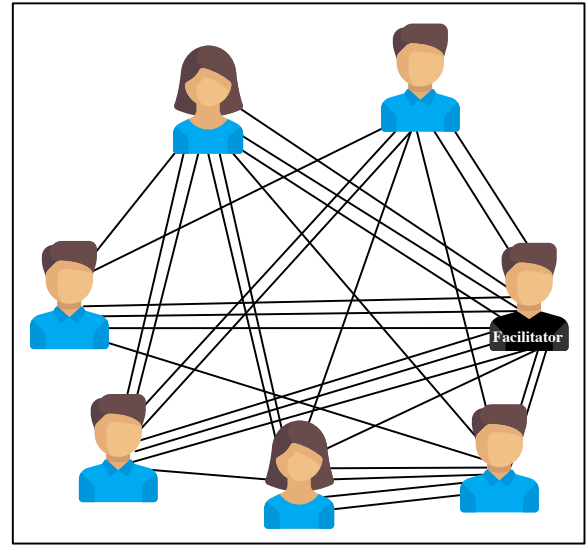

**Figure 2:** Conversation chart of a reporting meeting.

**Figure 3:** Conversation chart of a discussion meeting.

Facilitating the meeting can be even more challenging when the team members are distributed beyond the physical location of the meeting, which was the case in many of the teams we studied. The teams with facilitators who used video instead of only audio had more discussion meetings. One developer explained, *"Use of video changed our meetings because if you see people's reactions and their gestures, you immediately know whether they are listening or are bored, and whether they don't understand what you said."* Additionally, using a large video screen in the right height for standing helped team members acquire the feeling that everyone was a part of the same standing circle, which further increased the team spirit and improved the communication.

**When are the meetings held?**

Most agile teams conduct their daily stand-ups in the morning. A few team members told us that it is good to start the day by gathering the team to hear what people will be working on that day and whether this work will face any obstacles. Nevertheless, we found in many cases that an early start reduced the perceived value of the meeting and the effectiveness of the team. We observed that early meetings were often delayed a few minutes because some participants arrived late to work. Then the time spent in the meeting was more than the 15 minutes for those who showed up on time.

More importantly, many team members considered the meeting to be a disruptive interruption, particularly during programming activities. One developer said: *"I find the meeting useful, but the interruption it causes is disturbing. I have to detach myself from what I am working on, and when I get back it takes some time to start again where I left*

*off"*. Therefore, even when arriving early, some team members waited until after the meeting to start on challenging issues, and instead spent their time before the meeting on tasks not requiring concentration. This behaviour is natural as other research has found that most developers need more than 15 minutes to resume work after an interruption [12]. Starting early in the morning (e.g. 8.30 am) or late in the afternoon (e.g. 15.30 pm) was also seen as undermining the much-valued flexibility about work hours in the companies.

Participants said that starting later in the day (e.g., 10 am) was also problematic because the time between the daily meeting and lunch was too short to do anything useful. Starting right before lunch was perceived as a better option for many because then there was no added interruption and the team could go and eat lunch together.

**Who benefits most from the meetings?**

The way work was assigned and coordinated affected the dynamics of the stand-ups. In some teams, there were few interdependencies between tasks of different members. The main reason was that they had specialised roles and low knowledge redundancy; that is, team members had expertise in different technical areas and worked on separate modules. As a consequence, what one team member worked on was often uninteresting or irrelevant for other members than the Scrum Master or technical lead. The team members therefore usually did not pay attention when others were talking. When there is little interdependence among team members, there is less need for mutual adjustment and problem-focused communication, and as a result, most of the time is instead spent on reporting. Particularly, there is less interdependence in larger teams because everyone's work rarely depends on everyone else's work. Accordingly, we observed that larger teams had more reporting meetings. Norbank had the shortest meetings and the largest teams, partly as a consequence of having *BizDevOps* teams, which include people from business, development and operations. Problem-discussions among such a variety of roles would have required too much time and would not have been relevant for all.

The more people who attended a meeting, the less time was available for each person to be active, and so team members were generally less satisfied with large meetings. This is consistent with what we found in a recent survey of stand-ups, where there was a significant decrease in meeting satisfaction when the team comprised of more than 12 members [2].

Furthermore, we found that experience played an important role with respect to who benefitted from the meeting. Junior team members were generally more positive towards the meeting than seniors. Senior team members often perceived the daily stand-ups to give little personal value because they already knew what was going on in the team and did not receive any new information in the meeting.

Moreover, the positive attitude towards problem-solving discussions is more relevant to juniors because the problems they encounter are easier to solve in a daily stand-up. Senior team members often work on more complex tasks where the problems will require more expertise and time to resolve than a daily stand-up allows. Furthermore, when an experienced team member works on a complex task, there might be little progress for some days, and team members do not see repeating what is said in the previous meeting as valuable. One senior developer said: *"The intervals are too short to report on; often you feel that you did not have much progress from the day before"*.

We have also seen that senior team members attend more meetings, both internal to the team and external, in addition to the daily stand-ups, than junior team members. The daily meeting is then an additional daily interruption, which may reduce the well-being of team members, since a higher meeting load negatively affects the level of fatigue and subjective workload [11].

## Recommendations

In our longitudinal study, we first made a set of initial recommendations on the basis of our early interviews and observations. The recommendations were then tried out by several teams, which we then interviewed and observed. While there is no clear recipe for how to conduct a successful daily stand-up that fits all companies, we make the recommendations as follows based on all our data collection and analysis.

### Omit Question 1

To avoid the daily stand-up becoming a status-reporting meeting that does not realise its potential benefits, we suggest that team members do not report what has been accomplished since the last meeting (Q1). Eliminating this question will reduce status-reporting and self-justification, where participants explain in detail what they have done and why they have not achieved as much as expected. More time will then be available for discussing and solving problems. While it is valuable for team members to know the status and progress of the other team members' tasks, such information can be shared more efficiently by other means. For example, some teams display their status on a visual board, and some use a chatbot in Slack (an electronic communication tool) to collect status information from everyone, which is later posted to the whole team.

### Share facilitation responsibility

Facilitators need to be conscious about allotting equal time to all participants. It is easy (often unconsciously) to allow some people to talk more than others because the information from these people may be of particular interest to the facilitators. Furthermore, the facilitator must be aware of how the allocation of who speaks when during the meeting affects the conversation flow. Try out both an approach where the tasks on the board are discussed and a round-robin approach where the team members know without interruption who is to speak next.

One alternative to having one appointed facilitator is to circulate the facilitator role among a set of team members, which some of the observed teams did. When leadership is broadly distributed instead of centralized in the hands of a single individual, it is known as shared leadership. Shared leadership can empower a team and foster both task-related and social dimensions of a group's function, such as trust, cohesion and commitment. Since new facilitators may influence how the meeting is carried out, sharing the facilitator role will give the team opportunities to reflect upon different ways of conducting the daily stand-ups. If the facilitator is not the same person as the team leader, it is less likely that the daily stand-up becomes a status-reporting meeting.

**Meet right before lunch**
If a team decides to conduct daily stand-ups, the meetings must fit the rhythm of the team members' day and week. It is important to find the least disruptive time to reduce the potential for fragmented work. In many cases, conducting the meeting early in the morning has some drawbacks. Right before lunch may be a better time as it will merge two interruptions. Many teams tried it, and one respondent commented on not having the meeting in the morning: *"There are many advantages with the new time; one is that people have different preferences when it comes to what time to arrive at work. When having the stand-up late, all of us are able to have moments of flow before the meeting. It is also much easier to remember what you are working on. I know the trend has spread to other teams"*.

Furthermore, hungry team members are more prone to end the meeting on time, and in addition, team members are more likely to have lunch together, which will stimulate team cohesion. They can also continue discussions on their way to lunch, and during lunch, if needed.

**Adapt the frequency**
The daily stand-up meeting is, as the name indicates, intended to be conducted daily. However, we found that not all teams benefit from meeting five times a week. For example, small, collocated teams that have a high degree of informal communication may not need a daily meeting. Furthermore, if the team is large, there will be less need for mutual adjustment among all the team members. Therefore, it might be better to split the team into smaller teams and hold separate meetings for those who work together on a daily basis, and less frequent meetings for the whole team. Alternatively, team members may post answers to the three Scrum questions electronically on a daily basis, and then have a less frequent physical meeting where they can concentrate on the problem-focused communication.

**TABLE 3:** Recommendations for daily stand-up meetings

| |
|---|
| **Stop asking "What did you do yesterday?"** |
| • Reduce time spent on status-reporting and self-justification <br> • Focus on future work, particularly considering dependencies and obstacles <br> • Spend time discussing and solving problems as well as making quick decisions |
| **Optimize the communication pattern** |
| • Share the leadership to increase joint responsibility, for example by rotating the facilitator role <br> • Team members should communicate with each other, and not report to the facilitator |
| **Find the least disruptive time** |
| • Consider scheduling the meeting to right before lunch to decrease the number of interruptions |
| **Find the frequency that gives most value** |
| • In a well-communicating team, stop meeting daily if 3-4 times a week is sufficient <br> • In a large team, some team members do not need to meet as frequently as others, depending on the interdependencies among tasks and members |

## Conclusion

We have worked with many teams in different companies who have experimented with our recommendations (Table 3). They have stopped asking Q1: "What did you do yesterday?", started sharing the facilitation responsibility to optimize the communication pattern and changed the frequency and time of day of the meeting. In particular, having the meeting right before lunch has been widely adopted and appreciated. Furthermore, we have yet to meet any teams who have reinstated Q1 after having tried a period without it. Members appreciate focusing on the work that is ahead of them and being allowed to discuss problems and solutions. The one recommendation that is less adopted is sharing the facilitator role, probably because Scrum Masters and team leaders see it as their responsibility to facilitate the meetings.

When we help companies, we always advise that they collect information about the daily meeting to gain insight into their teamwork and thus how the productivity of the team can be enhanced. Having a person from another team draw a conversation chart (see Figure 2) is a technique that may provide insight as well as a valuable basis for discussions about daily stand-ups. Relevant questions to discuss are: How can all team members benefit from the meeting? Are people talking to each other or to the facilitator? Is it reporting or discussion? Who should facilitate the meeting? Is every team member talking approximately the same amount of time? Should the meeting focus on people or tasks?

We advise all agile teams to try to improve their daily stand-up meetings by experimenting with breaking the rules that they follow. Taking responsibility for improving the meetings will make the team more self-managed. An agile software team should continuously inspect and adapt the daily stand-up meeting to fulfil the team's needs. Some of the adjustments might be about challenging the old mindset, such as conducting the meetings daily and in the morning, and discussing only the three Scrum questions. Teams that make appropriate adjustments become more productive.

## References

[1] VersionOne, "VersionOne 12th Annual State of Agile Report," Mar. 2018, pp. 1–16.
[2] V. Stray, N. B. Moe, and G. R. Bergersen, "Are Daily Stand-up Meetings Valuable? A Survey of Developers in Software Teams," in *Agile Processes in Software Engineering and Extreme Programming*, vol. 283, no. 20, Cham: Springer International Publishing, 2017, pp. 274–281, https://doi.org/10.1007/978-3-319-57633-6_20
[3] H. Mintzberg, *Mintzberg on management: inside our strange world of organizations.* Free Press, 1989.
[4] J. A. Allen, N. Lehmann-Willenbrock, and S. J. Sands, "Meetings as a positive boost? How and when meeting satisfaction impacts employee empowerment," *Journal of Business Research*, vol. 69, no. 10, Oct. 2016, pp. 4340–4347, https://doi.org/10.1016/j.jbusres.2016.04.011
[5] K. Schwaber and J. Sutherland "The Scrum Guide", http://scrumguides.org/docs/scrumguide/v2017/2017-Scrum-Guide-US.pdf, Nov. 2017, pp. 1-19.
[6] S. Kauffeld and N. Lehmann-Willenbrock, "Meetings Matter: Effects of Team Meetings on Team and Organizational Success," *Small Group Research*, vol. 43, no. 2, Mar. 2012, pp. 130–158.
[7] V. Stray, D. I.K. Sjøberg, and T. Dybå, "The daily stand-up meeting: A grounded theory study," *Journal of Systems and Software*, vol. 114, Apr. 2016. pp. 101–124, https://doi.org/10.1016/j.jss.2016.01.004
[8] H. Nyrud and V. Stray, "Inter-team coordination mechanisms in large-scale agile," XP2017 Scientific Workshops, ACM, 2017, pp. 1–6. https://doi.org/10.1145/3120459.3120476
[9] G. Klein, "Naturalistic Decision Making," *Human Factors: The Journal of the Human Factors and Ergonomics Society*, vol. 50, no. 3, Jun. 2008, pp. 456–460, https://doi.org/10.1518/001872008X288385
[10] C. Duhigg, "What Google Learned From Its Quest to Build the Perfect Team," *The New York Times*, pp. 1–18, 25-Feb-2016.
[11] A. Luong and S. G. Rogelberg, "Meetings and More Meetings: The Relationship Between Meeting Load and the Daily Well-Being of Employees.," *Group Dynamics: Theory, Research, and Practice*, vol. 9, no. 1, 2005, pp. 58–67, http://dx.doi.org/10.1037/1089-2699.9.1.58
[12] C. Parnin and S. Rugaber, "Resumption strategies for interrupted programming tasks," *Software Quality Journal*, vol. 19, no. 1, Aug. 2011, pp. 5–34, https://doi.org/10.1109/ICPC.2009.5090030

## Tweets

- Daily stand-ups – what, how, when and for whom? …read more @IEEESoftware
- Recommendations on how to make #agile teams more effective by adjusting the most popular practice @IEEESoftware
- Who benefits most and what is discussed in daily stand-up meetings? Find out now… @IEEESoftware
- Adjust the daily stand-up meeting to empower your team. Read about main benefits and problems of the meeting @IEEESoftware
- Don't start the day with daily stand-up meetings…read more @IEEESoftware

## About the authors

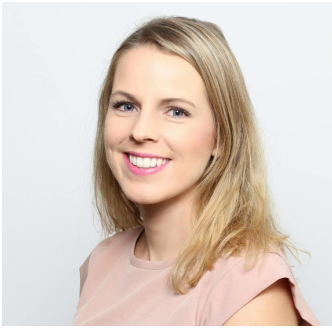

**Viktoria Stray** is an associate professor at the University of Oslo's Department of Informatics. Her research interests include agile methods, global software engineering, teamwork, coordination and large-scale development. Her main focus is to improve productivity in software projects. Stray has a PhD in software engineering and has worked several years in the software development industry. She also holds a research position at SINTEF. She's a member of IEEE and ACM. Contact her at stray@ifi.uio.no.

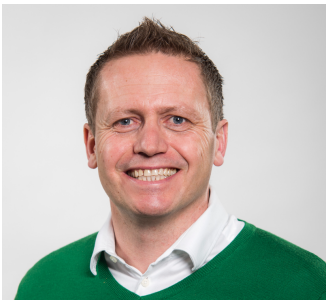

**Nils Brede Moe** is a senior scientist at SINTEF. He works with software process improvement, intellectual capital, and agile and global software development. His research interests are related to organizational, socio-technical, and global/distributed aspects. His publications include several longitudinal studies on self-management, decision making, innovation, and teamwork. He holds an adjunct position at the Blekinge Institute of Technology in Sweden. Contact him at nilsm@sintef.no.

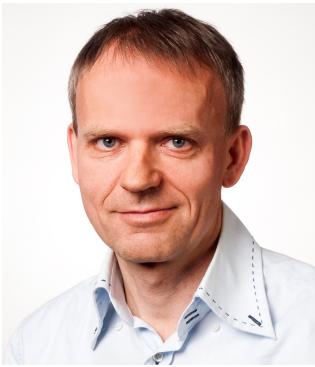

**Dag I.K. Sjøberg** is professor of software engineering at the University of Oslo. His main interests are the software life cycle, including agile and lean development processes, and empirical research methods in software engineering. Sjøberg has a PhD in computing science from the University of Glasgow. He's a member of IEEE and ACM. Contact him at dagsj@ifi.uio.no.